\documentclass[fleqn,10pt]{wlscirep}
\usepackage[utf8]{inputenc}
\usepackage[T1]{fontenc}
\usepackage{lineno}

\title{Echo chamber formation sharpened by priority users}

\author[1,*]{Henrique F. de Arruda}
\author[1]{Kleber A. Oliveira}
\author[1,2,3]{Yamir Moreno}
    
\affil[1]{CENTAI Institute, Turin, Italy}
\affil[2]{Institute for Biocomputation and Physics of Complex Systems (BIFI), University of Zaragoza, Zaragoza, Spain} 
\affil[3]{Department of Theoretical Physics, Faculty of Sciences, University of Zaragoza, Zaragoza, Spain}

\affil[*]{henrique.f.arruda@centai.eu}

\begin{abstract}
Priority users (e.g., verified profiles on Twitter) are social media users whose content is promoted by recommendation algorithms. However, the impact of this heterogeneous user influence on opinion dynamics, such as polarization phenomena, is unknown. We conduct a computational mechanistic investigation of such consequences in a stylized setting. First, we allow priority users, whose content has greater reach (similar to algorithmic boosting), into an opinion model on adaptive networks. Then, to exploit this gain in influence, we incorporate stubborn user behavior, i.e., zealot users who remain committed to opinions throughout the dynamics. Using a novel measure of echo chamber formation, we find that prioritizing users can inadvertently reduce polarization if they post according to the same rule but sharpen echo chamber formation if they behave heterogeneously. Moreover, we show that a minority of extremist ideologues (i.e., users who are both stubborn and priority) can push the system into a transition from consensus to polarization with echo chambers. Our findings imply that the implementation of the platform's prioritization policy should be carefully monitored in order to ensure there is no abuse of users with extra influence.
\end{abstract}


\begin{document}
\flushbottom
\maketitle
\section{Introduction}

After undergoing a recent shift in its business model, Twitter, a well-studied social media platform~\cite{karami2020twitter}, has introduced substantial revisions to its verification policy. Users labeled as verified by the platform went from being ``active, notable, and authentic''~\cite{twitter_old_verified} to those who meet substantially more permissive criteria~\cite{twitter_new_verified} and pay a monthly fee to a subscription plan. The benefits of the subscription involve forms of algorithmic boosting, such as prioritized ranking in replies and searches. However, inadvertently promoting these users may trigger unknown macroscopic effects in the system, especially regarding opinion dynamics~\cite{sirbu2019algorithmic}. Given this problem, our paper addresses the following research question: \textit{what is the effect of inadvertent social media content promotion on opinion polarization and the formation of echo chambers?}

Social media platforms implemented verification processes to label some users differently than the rest. On Facebook~\cite{facebook_verified} and Instagram~\cite{instagram_verified}, it means the platform itself vouches for the authenticity of the associated user account. Concurrently, the role influential individuals play in the spread of information and opinion has been known from much before social media platforms~\cite{watts2007influential,centola_pescosolido_smith_2021}. Hence, many people in society who are famous or influential regardless of social media tend to rely on this feature so that their online influence is not misused or mischaracterized. In prominent cases, a verified account of an influential individual is an active public speech stage where they can perform and carry marketing or political campaigns~\cite{ott2017trump}.

The rules for each verification procedure and benefits associated with verified accounts change according to the business model of the platform. As a consequence, there has been much debate~\cite{morris2012tweeting,indraneil2019elites,vaidya2019verified,alkhamees2021trustworthiness} about how these verified users exert more influence compared to the other users. In the case of Twitter, the change of verification policy has been a cause for concern for institutions that design policies involving situations affected by opinion polarization. For instance, the French Commission for Campaign Accounts and Political Financing (CNCCFP) decided not to allow candidates to subscribe to Twitter's new verification system ahead of the 2024 French elections~\cite{french_commission_elections}. They understand the subscription constitutes an increase of visibility and reach of the subscribed account and hence a form of sponsored advertising.

It is not surprising to see policymakers acting preemptively toward perturbation on already polarized scenarios. Opinion polarization is not an easy phenomenon to intervene or control, and as such has been increasingly studied because it is critical to public debate and collective decisions~\cite{Bak-Coleman2021stewardship, lorenz-spreen2020behavioural}. Many researchers turned their attention to studying this phenomenon ever since measurement via social media data~\cite{delVicario2016misinformation, brady2017emotion} become available, particularly in significant periods (such as upcoming elections or referenda) of democracies around the world~\cite{narayanan2019news,baker2020elections,machado2018news,delvicario2017mapping}. Computational models of opinion dynamics have been developed to explain the formative mechanisms behind polarization~\cite{noorazar2020opiniondynamics}. Agent-based models are a class of such models that computationally simulate the system through numerous agents whose automated behavior is ore-specified by a set of rules. The majority of the agent-based models encode opinion mechanisms formed from the interaction among agents~\cite{deffuant2000mixing, degroot1974reaching,hk2002opinion,galam2002minority}. As a result, network descriptions are largely used in the field. For instance, the persuasion mechanism, characteristic of social networks, has been modeled in the Sznajd model~\cite{he2004sznajd, sznajd2021review}. Adaptive networks have also been modeled and thus identified as a mechanism for the formation of echo chambers~\cite{holme2006nonequilibrium, benatti2020opinion, baumann2020modeling}.

In tandem with the verification policy change, Twitter also started to limit its data access to researchers. To explore the implications of the platform alterations, we approach our research question computationally by incorporating heterogeneous content promotion into an agent-based simulation of opinion dynamics in a social media setting through different types of users. We modify a previously developed opinion model~\cite{de2022modelling} on adaptive networks to incorporate heterogeneous user behavior. The model was capable of reproducing many opinion configurations (from consensus to echo chambers) even with its previous setup (homogeneous user behavior). Valensise~\emph{et al.}~\cite{valensise2023drivers} conducted tests on similar mechanisms to verify which has better data support from different social media platforms. They found that the opinion model by Arruda~\textit{et al}, which we develop from, was better suited to reproduce polarization measurements from Twitter than other social media platforms, adding to the results of the original paper~\cite{de2022modelling}.

In this study, we investigate the effects of priority accounts on social media. We start by implementing priority users to the agent-based simulation and test how they influence macroscopic measures of polarization. After this, we incorporate a second novel type of user, named stubborn, through which we study how far can the gain of influence from prioritization be pushed towards perturbating the dynamics. To do so, we investigate content promotion of extremist or centrist users via the priority policy.

To adequately measure echo chamber formation, we propose a novel measure that incorporates the opinion density map into a single distribution. Our results reveal that the presence of extremist stubbornness, even in small proportions, can significantly drive opinion polarization and strengthen echo chambers. Moreover, we observe that stubborn centrists, when non-priority, can reinforce polarization in the network. These findings highlight the potential risks of policy changes in social media and emphasize the complex interplay between network structure, user behavior, and opinion dynamics in social media environments. Overall, our study elucidates which mechanisms are enough to drive macroscopic changes to the dynamics and, by doing so, provides insights into understanding the role of content promotion policies in shaping public discourse and opinion polarization in online media. Finally, we emphasize that the change in verification policy also raises research questions about the dynamics of misinformation. However, in this paper, we limit the scope of our investigation to mechanisms of opinion polarization because empirical data are now limited after the change in the data access policy.

\section{Results}

\subsection{Background and Model} \label{sec:model_description}

Our model is based on a modification of the one introduced by de Arruda~\textit{et al.}~\cite{de2022modelling}. The initial setup is a directed network where nodes represent social media users, each with a continuous opinion bounded between -1 and 1. A user points to another if they receive content from them (followership relation); that is, the content is spread following the opposite direction of edges. Furthermore, the content has its own opinion value $\theta$, which is always fixed and uniformly random.

Each iteration is summarized with the following steps.

\begin{enumerate}
    \item \textit{\textbf{Activating:}} a node $i$ is uniformly randomly picked to become active. 
    \item \textit{\textbf{Posting:}} The active node $i$ posts a content of opinion $\theta$ with probability defined by a rule comparing its opinion $b_i$ with $\theta$.
    \item \textit{\textbf{Receiving:}} Each follower $j$ of $i$ receives the post or not according to another probabilistic rule, also a function of their opinion $b_j$ and $b_i$. We say there is a recommendation algorithm controlling this particular rule.
    \item \textit{\textbf{Realigning:}} Those who received the content may either move their opinion towards or away from $\theta$ by a constant $\Delta = 0.1$. They are repulsed away from the post with probability $|\theta - b_j|/2$ and attracted otherwise.
    \item \textit{\textbf{Rewiring:}} A stochastic bounded-confidence edge condition is checked for each follower $j$, in which the bigger the difference between $b_j$ and $b_i$ is, the more likely the edge from $j$ is pointed to another random user instead of $i$.
\end{enumerate}

In this work, we modify this model by introducing two additional types of users, each with their own behavior.~A \textit{priority user} posts content that is prioritized by the recommendation algorithm, always being received by followers.~A \textit{stubborn user} is a user that never changes their opinion.~Notice these two types may overlap; when they do, we call such user an \textit{ideologue}. Users who do not belong to any of those types are called \textit{normal users}. For convenience, we also refer to a user with opinion at either 1 or -1 as \textit{extremist}, and another with opinion 0 as \textit{centrist}.

For the sake of disambiguation, we note that naming users with the two simultaneous behaviors as ideologues do not directly imply they possess any topological advantage in terms of the network structure (e.g., increased number of followers). We only postulate these users have priority in terms of the recommendation algorithm and that they are committed to a point of view in the form of stubborn behavior. We also show in the later sections that a small number of users with those two behavior modifications is enough to cause strong changes in the system. Moreover, for some configurations, we can observe an emergent core-periphery structure, explained in Section S3 of Supplementary Material.

In terms of the implementation of these new features, we refer to the base model from de Arruda~\textit{et al}.~\cite{de2022modelling} and its parameter space. The probabilistic rules in each step work as filters to check whether all the steps down to the edge rewiring happen.

In our extension, we choose two probabilistic rules for the posting step. The first one is calculated as
\begin{equation}
    P^{\text{con}}_p( \theta, b_i) = \cos^2\left( \frac{\pi}{2} |\theta - b_i| \right).
    \label{eq:P_p_con}
\end{equation}
\noindent Users who follow this rule do \textit{controversial posting}, since they may send content both aligned and opposed with their own opinion. All normal users follow this rule.

Stubborn users use an alternative rule we name \textit{aligned posting}, defined by
\begin{equation}
    P^{\text{ali}}_p( \theta, b_i )= 
\begin{cases}
    \cos^2\left( \frac{\pi}{2} |\theta - b_i| \right), & \text{if } |\theta - b_i| \leq 1\\
    0,              & \text{otherwise}.
    \label{eq:P_p_ali}
\end{cases}
\end{equation}

\noindent This rule means they send content close to their opinion so as to represent a stylized zealot behavior.

As per the priority users, the receiving step associated with their posting is modified. When the user $i$ is not a priority user, the social network recommendation algorithm modulates how each follower $j$ receives content from $i$ through the probabilistic rule
\begin{equation}
    P_r(b_i, b_j, \phi) = \cos^2\left( \frac{\pi}{2} |b_i - b_j| + \phi \right),
    \label{eq:rp_I}
\end{equation}
where the parameter $\phi$ controls the starting point of the cosine-squared function.
However, if the active user $i$ is a priority user, the recommendation algorithm overrides this step, and the content is automatically received by all followers.

All other steps are implemented the same from the base model~\cite{de2022modelling}, with the difference that now we consider directed edges (out-neighbors do not receive content). Thus, when a user followed by stubborn users becomes active, the stubborn followers are still checked for rewiring even after skipping the realigning step because the active user may have had their opinion changed from other iterations.

By controlling the function defined by Eq.~\ref{eq:rp_I}, the parameter $\phi$ can poise the base model at polarization or consensus. This relation is further described in Fig.~S1 in the Supplementary Material. As a reference, we pick $\phi=0.785$ to test the model at polarization and $\phi=1.473$ to test it at consensus.

\subsection{Homophilic bimodality coefficient}
While the polarization of the system is adequately described by the distribution of opinions $b$, it may happen through different network configurations. A polarized distribution $b$ may be associated with a highly divided network, where intra-group opinion is aligned but inter-group is opposite (echo chamber), but it can also be found in a more homogeneously connected network, where many edges tie together people with opposite opinions.

Cota~\emph{et al.}~\cite{cota2019quantifying} proposed a method based on plotting a density map to quantify the effect of separate groups with different opinions in a social network. In this approach, the distribution of opinions $b$ is plotted against the average opinion of each user's outgoing neighbors, denoted as $b_{NN}$. That is, the density map captures the relationship between a user's opinion and the opinions of the users they follow. When the network is divided into internally aligned groups that oppose each other externally, there is a higher density in the first and third quadrants of the associated map. This observation makes it possible to quantify the echo chamber effect within the network. The density map approach has been widely used in various studies~\cite{baumann2020modeling,cinelli2020echo,cota2019quantifying,de2022modelling,valensise2023drivers,hohmann2023euclidean}.

Arruda~\textit{et. al} employed an auxiliary measure called the bimodality coefficient $BC$ that was used to assess the degree of bimodality in the distribution of opinions. Given a sample distribution $b$ of size $n$ (the number of users), the bimodality coefficient $BC(b)$~\cite{pfister2013good} is defined as 
\begin{equation}
    BC(b) = \frac{g^2 + 1}{k + \frac{3(n-1)^2}{(n-2)(n-3)}},
    \label{eq:BC}
\end{equation}
where $g$ and $k$ are the \emph{sample skewness}~\cite{kokoska2000crc} and \emph{excess kurtosis}~\cite{kokoska2000crc} of the distribution $b$, respectively. Empirically, if $BC(b) > 5/9$, the distribution is considered bimodal~\cite{pfister2013good}. This measure only considers bimodality from $b$ and not from the density map with both $b$ and $b_{NN}$.

Martin-Gutierrez~\emph{et al.}~\cite{martin2023multipolar} projected the opinions through a principal component analysis to find the best multipolar projection. However, since we are considering a single opinion, we would not be able to distinguish between polarization with and without the formation of echo chambers. Here, we propose a novel measure to enhance the neighborhood aspect and capture echo chamber formation as the bimodality in the combined dimensions of $b$ and $b_{NN}$. Firstly, we rotate the $b \times b_{NN}$ map by $45^{\circ}$ using a rotation matrix $R$. By applying this rotation, we align the axes along the major and minor axes of the distribution, effectively combining the two dimensions. Subsequently, we calculate $BC$ using only the first dimension of this rotated map. This novel measure, which we refer to as the homophilic bimodality coefficient $BC_{\text{hom}}$, aims to assess the bimodality of both $b$ and $b_{NN}$. 

Following the previous notation, a distribution $b^\dagger$ from the projected diagonal is obtained from
\begin{equation}
    b^\dagger = [ [b, b_{NN}] R ]_{i 1} \quad \text{for} \quad 1 \leq i \leq n,
\end{equation}
where $[b, b_{NN}]$ is of order $n{\times}2$ and $R$ is the $2{\times}2$ rotation matrix of $45^{\circ}$.
Thus, the homophilic bimodality coefficient $BC_{\text{hom}} (b, b_{NN})$ of the original distribution $b$ associated with a $b_{NN}$ is
\begin{equation}
    BC_{\text{hom}}(b, b_{NN}) = BC(b^\dagger).
    \label{eq:BC_hom}
\end{equation}
In other words, the transformation combines both the overall opinion distribution in the system and the neighborhood alignment into a single distribution, which allows us to clarify whether the opinion poles also coincide with aligned networked clusters through the bimodality of the transformed distribution. Moreover, we often employ our measurements to understand what happens to the opinions of other users instead of stubborn (who have their opinion set at initialization) or priority. Therefore, for convenience, we define $\hat{b}$ as the distribution opinion of only normal users. Analogously, the corresponding $\hat{b}_{NN}$ is the distribution of average neighbors' opinions only from normal users, but considering all neighbors.

\begin{figure*}[!th]
  \centering
    \includegraphics[width=1\textwidth]{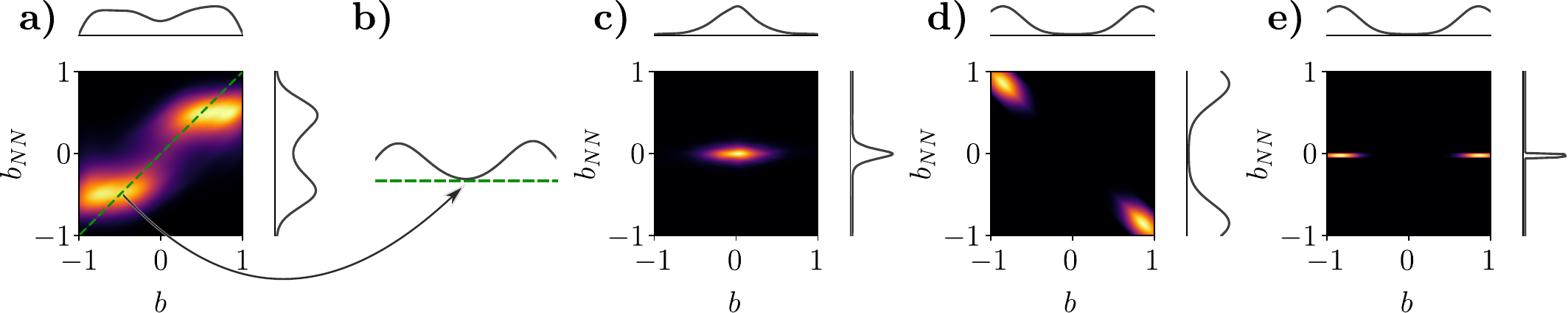}
  \caption{\textbf{Measuring echo chambers with $BC_{\text{hom}}$:} Panel (a) shows the density map of $b$ against $b_{NN}$ and panel (b) shows the projected diagonal of the density map from (a). For this example, $BC(b)= 0.583$ and $BC_{\text{hom}}(b, b_{NN})=0.667$, which shows how the measure $BC_{\text{hom}}$ enhances the echo chamber detection by encoding it into the bimodality of the rotated distribution. A consensus scenario (panel (c)) has unimodal distributions in both axes, which should not be significantly changed by the rotation ($BC(b)=0.237$ and $BC_{\text{hom}}=0.226$). Panel (d) represents a scenario of conflict, which is polarized and divided, but its groups do not have aligned opinions; hence our measure penalizes it ($BC(b)=0.961$ and $BC_{\text{hom}}=0.415$). In the last case, in panel (e), we show a polarized scenario with no clear division in the network. In this case, $BC(b)=0.961$ and $BC_{\text{hom}}=0.926$. This is a case in which $BC_{\text{hom}}$ does not inform better whether echo chambers formed in the system.}
  \label{fig:HBC}
\end{figure*}

To illustrate the effectiveness of this measure, we simulate a polarized base scenario ($\phi=0.785$) in a randomly wired network of size $n=10^4$ and mean in-degree $z=8$, where nodes are initialized with a randomly uniform opinion between -1 and 1. Fig.~\ref{fig:HBC}(a) shows the density map of $b$ against $b_{NN}$ obtained through this base polarized scenario, and Fig.~\ref{fig:HBC}(b) exhibits the corresponding transformed distribution $b^\dagger$ described above. The other panels explain how the measure reacts to other scenarios, such as consensus (\ref{fig:HBC}(c)) and other types of polarization.

We notice that when the homophilic bimodality coefficient $BC_{\text{hom}}$ exceeds the $5 / 9$ threshold, it does not mean the system is organized into exactly two groups, one with positive opinions and the other with negative ones. As pointed out by Hohnmann~\emph{et al.}~\cite{hohmann2023euclidean}, echo chambers may happen in a network divided into several groups. It may well be the case that a high $BC_{\text{hom}}$ points to the presence of several network clusters following the pattern of aligned intra-group opinion and opposing inter-group opinion.

\subsection{Effect of priority users}
\label{sec:verified_accounts}
We now systematically test what happens to the system when the dynamics are initialized with an increasing fraction of priority users. The stubborn behavior is considered in the following sections. The initial configuration is the base case, shown in Fig.~\ref{fig:HBC}(a). More details on how we define this initial state are described in Section~\ref{sec:methods}.

Regarding the network structure, we present results for homogeneous networks, generated with the Erd\H{o}s-R\'enyi~\cite{erdos1960random} (ER) model, and compare what we obtain from the different posting functions, $P^{\text{con}}_p$ and $P^{\text{ali}}_p$. We test different network structures (e.g., networks with communities and scale-free degree distributions), but the results are similar to those presented here. Therefore, these results are not shown.

\begin{figure}[!th]
  \centering
    \includegraphics[width=0.499\textwidth]{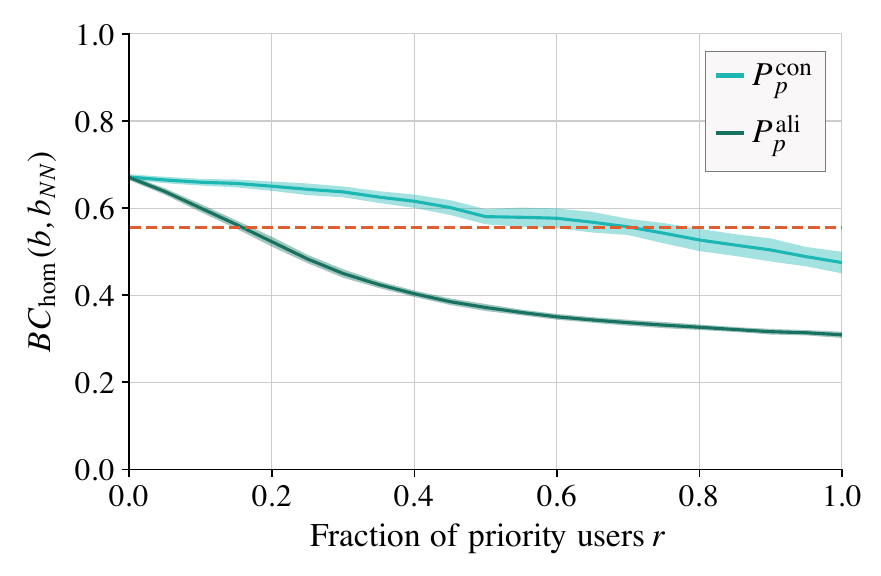}
  \caption{\textbf{Effects of priority users with different posting behaviors:} the measure $BC_{\text{hom}}$ is used to compare two cases of posting functions of the priority users. When there is no priority user, the parametrization recovers the case shown in Fig.~\ref{fig:HBC}a, meaning the system is parameterized at echo chamber formation. Priority users doing aligned posting drive the opinion distribution quickly out of polarization as they increase, but not when they do controversial posting for most of the range shown. The solid lines represent an average of over 100 executions of the dynamics, and the shaded regions are the corresponding standard deviations.}
  \label{fig:BCXpercentage}
\end{figure}

Specifically, we test what happens to the system as priority users post contents with (i) $P^{\text{con}}_p$ or post contents with (ii) $P^{\text{ali}}_p$. The latter represents users who want to be verified in order to spread their opinions. Fig.~\ref{fig:BCXpercentage} shows $BC_{\text{hom}}$ for both scenarios according to the fraction of users who are prioritized. In both cases, increasing the number of priority users decreases $BC_{\text{hom}}$, thus decreasing polarization. Furthermore, this effect is observed to be stronger when priority users do aligned posting, suggesting that polarization can be reduced if users avoid sending controversial content.

From the intuition gained from Fig.~\ref{fig:BCXpercentage} that the system moves away from polarization as the number of priority users doing aligned posting increases, we now build another step into our model. Up to now, we did not consider stubborn users. In order to understand how a fraction of priority users perturb the system with their influence, we observe what happens when these users are committed to opinions throughout the dynamics through the stubborn user behavior.

\subsection{How extremists can take advantage of being prioritized}
\label{sec:extremists_advantage}
Here we investigate how extreme opinions influence the system when set on priority users. We set them up as stubborn users with the posting function $P_p^{ali}$ to test the hypothesis that ideologues with extreme opinions can take advantage of being prioritized. Thus, their post contents are aligned with their own opinions. We set half of these users to opinion 1 and the other half to opinion -1 to understand the effect of groups of users with divergent opinions. Using this user behavior, we compare two scenarios, with and without priority accounts, in which the first type consists of ideologues. We calculate $BC_{\text{hom}}$ considering only the non-stubborn users to measure the impact of stubborn users on the network. This result is shown in Fig.~S2 of Supplementary Information~S2. Contrary to the results shown in Section~\ref{sec:verified_accounts}, for both cases, increasing the percentage of verified users strengthens their echo chambers. Furthermore, the verified case turned out to form stronger echo chambers for all the values tested, which shows that when verified users are extremists, polarization and division tend to increase faster.

\begin{figure}[!th]
  \centering
{\includegraphics[width=1\textwidth]{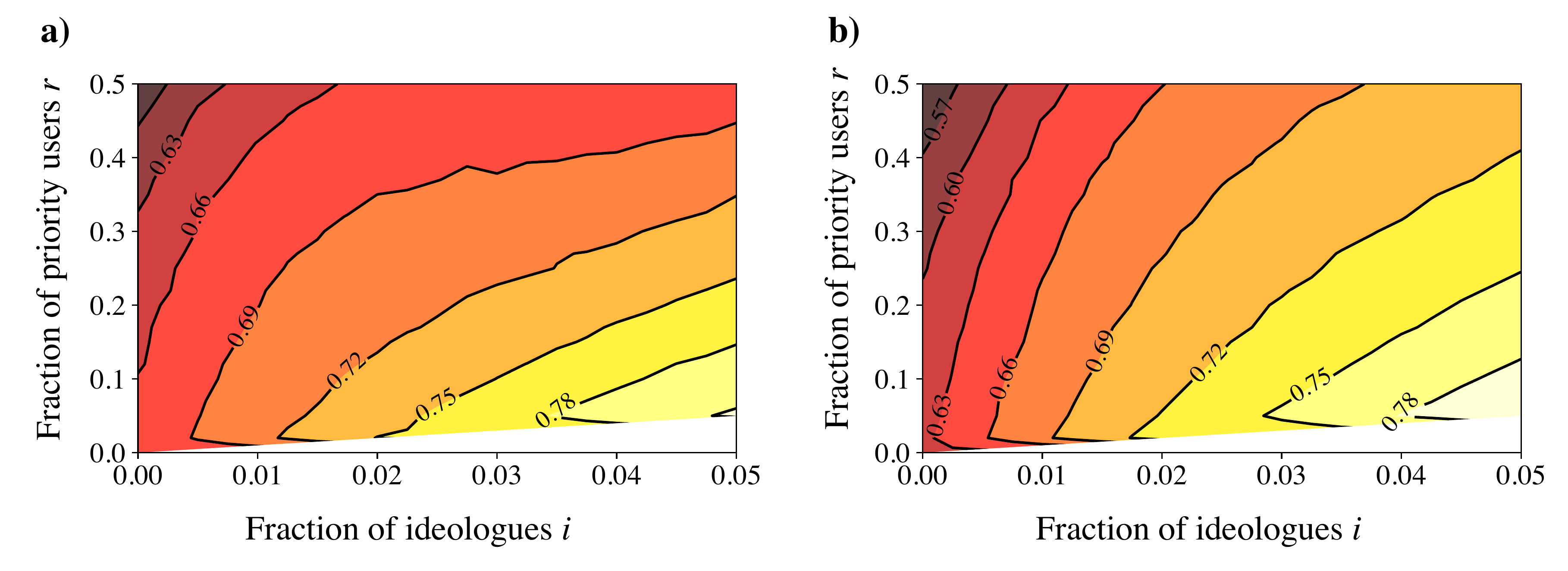}}
  \caption{\textbf{Effect of priority and ideologue users on echo chamber straightening:} Contour maps of $BC_{\text{hom}}(\hat{b}, \hat{b}_{NN})$ by varying the percentages of ideologues (both stubborn and priority) and priority users. While both axes represent fractions of the total number of users $N$, the stubborn users are randomly selected among the priority users to become ideologues (hence the white space for when there are more stubborn than priority users). Panels (a) and (b) represent the dynamics executed on random homogeneous and heterogeneous networks, respectively. The plots represent an average of 100 runs of the dynamics, and for all tests performed to plot these numbers, the standard deviation was less than 0.02.}
  \label{fig:Stubborn_verified}
\end{figure}

To have a more realistic comparison, for the next test, we map $BC_{\text{hom}}$ according to the percentages of priority and stubborn users. To do this, we vary the percentages of priority and stubborn users. Note that the fraction of stubborn is always less than or equal to the percentage of priority users because, in this case, all stubborn are also priority users. For this case example, we fix the posting as $P_p^{\text{ali}}$, which makes all the stubborn users behave as ideologues. Fig.~\ref{fig:Stubborn_verified} shows a contour map of this comparison. In Fig.~\ref{fig:Stubborn_verified}(a), we keep the example of the homogeneous random network. As expected, as the number of priority users increases, $BC_{\text{hom}}$ tends to decrease. But with some stubbornness, $BC_{\text{hom}}$ increases, showing that a few percent of stubbornness can make a difference.

We also test another network structure consisting of a network with a heterogeneous, scale-free (SF) degree distribution. 
The result obtained with this structure is shown in Fig.~\ref{fig:Stubborn_verified}(b). For SF, the effect of stubbornness is stronger for higher percentages of priority users than for the homogeneous network. Without stubbornness, the echo chambers are less well-defined. However, a smaller number of stubborn users can change the outcome of the dynamics and divide the network.

For the sake of completeness, in Supplementary Material~S3, we also mapped $BC_{\text{hom}}$ according to the percentage of priority and stubborn users, but in this case, we preserved posting as $P_p^{\text{con}}$. As a result, we find that when the users are stubborn but not ideologues, the percentage of priority users who are ideologues is the most important aspect in amplifying the echo chambers. It is worth noting that we tested a similar network configuration but with the in-degree and out-degree following a power law and a Poisson distribution, respectively. However, the results were found to be similar to the case of homogeneous random case. This is because our opinion dynamics change of the out-edges due to the rewirings. 

\begin{figure}[!th]
  \centering
  \includegraphics[width=1\textwidth]{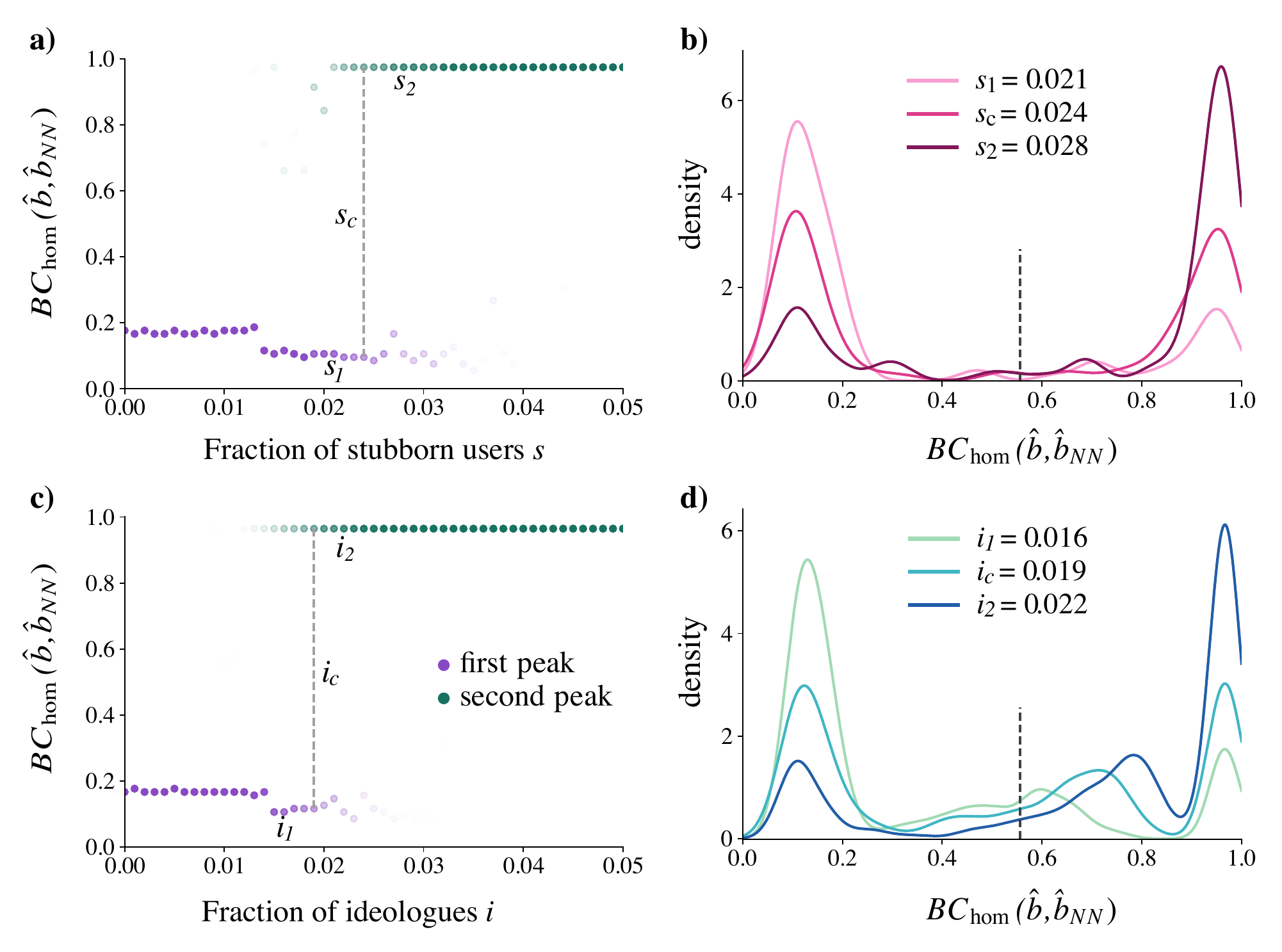}
  \caption{\textbf{Echo chamber formation accelerated by extremist ideologues:} the system undergoes a transition from consensus to polarization with echo chambers captured by the measure $BC_\text{hom}$ as the fraction of extremist users increases. In panel (a), the extremist users are only stubborn and in panel (c), they are ideologues (both stubborn and priority). In those panels, values of $s$ and $i$ are associated with the two peaks of the distribution of $BC_\text{hom}(\hat{b}, \hat{b}_{NN})$ values from 500 simulations. These peaks are obtained by dividing the 500 values around the threshold $5/9$. The transparency of points corresponds to how many simulations fall into each side. The purple dashed line is drawn for $s_c=0.024$ in panel (a) and $i_c=0.019$ in panel (c), where the threshold of bimodality divides the distribution of $BC_\text{hom}$ into two sides with approximately equivalent mass. That is, at $i_c$ and $s_c$, echo chambers are likely to be formed or not with similar odds. In panels (b) and (d), we show samples of those distributions for three values of $s$ and three values of $i$, respectively, including $s_c$ and $i_c$. The dark dashed vertical line marks the $5/9$ threshold.}
  \label{fig:extremists_transition}
\end{figure}

Next, we study the simulation outcomes in a consensus scenario, the base case set with $\phi = 1.473$. In particular, we gradually increase the fraction of extremist users (i.e., with opinion 1 or -1) who are stubborn while monitoring the $BC_{\text{hom}}$ measure as an order parameter, which is shown in Fig.~\ref{fig:extremists_transition}(a). Counterintuitively, a small minority of such users is able to drive the system from consensus to polarization with echo chambers, much like in the fashion of a social norm change via critical mass~\cite{centola2018tipping}. We then proceed to show how this transition happens even faster (at a fraction about $20 \%$ smaller) when this minority is also made of priority users in Fig.~\ref{fig:extremists_transition}(c). While the transition looks smoother in the case these users are only stubborn (Fig.~\ref{fig:extremists_transition}(b)) for our choice of the order parameter ($BC_{\text{hom}}$), it has a richer behavior for when they are ideologues (both stubborn and priority) as per observed in the mass emerging right before the second peak of the distributions (Fig.~\ref{fig:extremists_transition}(d)).

\subsection{Opinion polarization can still happen with centrist users}
For the sake of completeness, we also analyze the scenario in which the stubborn users are centrists. To do so, we set them with opinions equal to zero. As well as the simulations presented in Section~\ref{sec:extremists_advantage}, these users tend to post content similar to their opinions. 

It should be noticed that the measure $BC_{\text{hom}}$ was designed to accentuate the bimodality present in both the overall opinion distribution $b$ and the opinion alignment $b_{NN}$ by combining them into a single distribution. In the case either of these is not bimodal prior to the rotation, $BC_{\text{hom}}$ is not reliable to tell whether users connect with aligned neighbors. On the other hand, the angle of $45^{\circ}$ is chosen to punish cases in which both $b$ and $b_{NN}$ are bimodal, but users are connected with neighbors of different opinions. In that way, the resulting $b^\dagger$ from the rotation will likely return a value lower than the threshold $5/9$ (see, for instance, Fig.~\ref{fig:HBC}(e)). 
Since $b_{NN}$ is not bimodal for all results shown in this section, we analyzed the following results considering only $BC$. 

\begin{figure}[!th]
  \centering
{\includegraphics[width=1\textwidth]{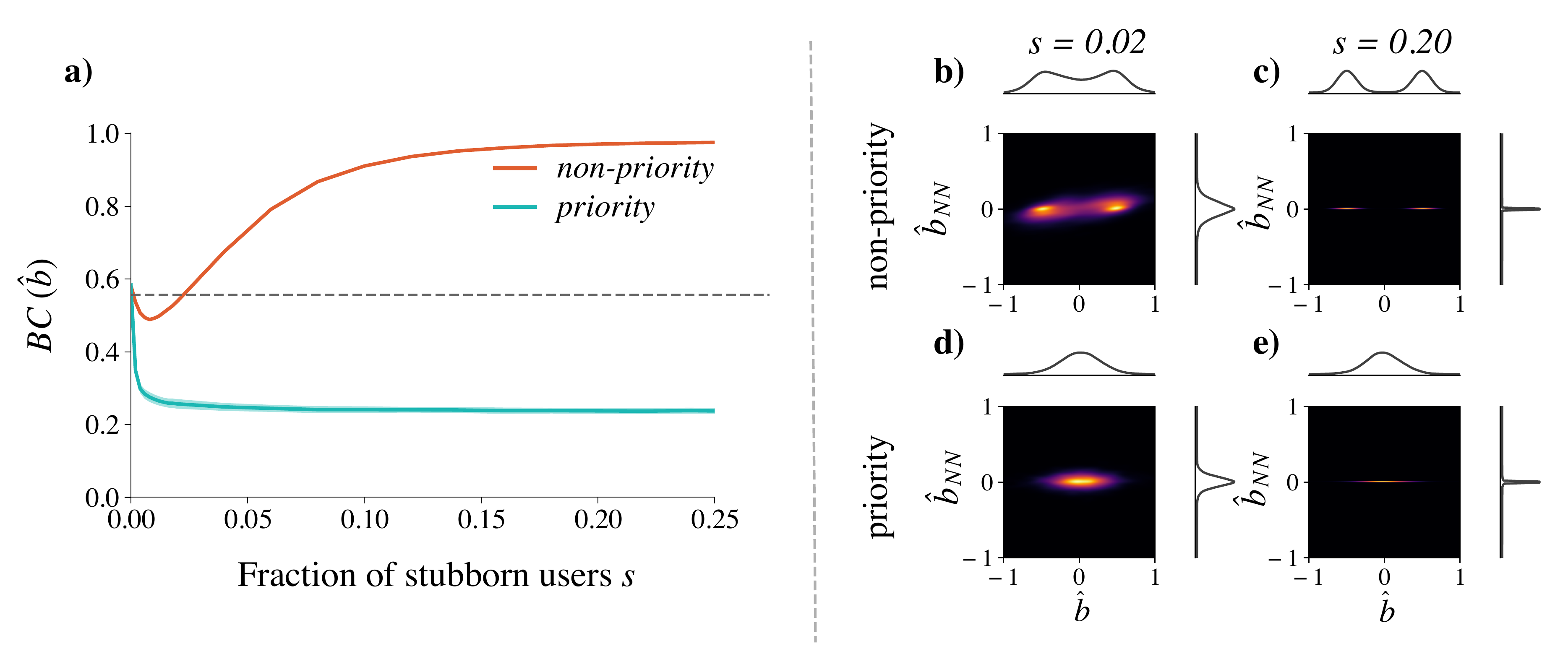}}
  \caption{\textbf{Impact of centrists on network polarization:} Panel (a) shows the comparison of the opinion polarization of ideologues and non-ideologues over the fraction of stubborn users. Panels (b) - (e) show the $\hat{b} \times \hat{b}_{NN}$ for a single run of the dynamics, for different fractions of stubborn users, and for priority and non-priority users. Even though both configurations start from a polarized point ($BC > 5/9$), stubborn centrists are not able to sustain a depolarization of the system, as opposed to centrist ideologues (who are both stubborn and priority).}
  \label{fig:comparison_priority_non-priority_centrists}
\end{figure}

We compare priority and non-priority users by varying the fraction of stubborn users (see Fig.~\ref{fig:comparison_priority_non-priority_centrists}). In general, the results are different when we contrast both scenarios. With a small fraction of stubborn users, the opinions tend to be strongly polarized in the case in which there are no non-priority users, while priority users (i.e., ideologues) induce the consensus. In the case of non-priority, for a fraction of stubborn users close to $0.6$, the opinion distribution tends to be unimodal, but for slightly higher values, it tends to be bimodal (see Fig.~\ref{fig:comparison_priority_non-priority_centrists}(b)). Furthermore, for higher values of the fraction of stubborn users, $\hat{b}$ is significantly bimodal (see Fig.~\ref{fig:comparison_priority_non-priority_centrists}(c)). In the case of priority users, as the fraction of stubborn increases, $BC(\hat{b})$ decreases until the scenario of consensus but maintains a diverse range of opinions. 

In all cases, there are no echo chambers because of a mechanistic implication of the rewiring rule in our model. That is, centrist stubborn users have probability 0 of losing followers, and may only gain them over time. After a sufficient number of iterations, most of the normal users are not connected to each other but only to the centrist stubborn users, which is akin to a hub-to-spoke core-periphery structure~\cite{gallagher2021coreperiphery}. This effect is visible in panels (b) to (e) of Fig.~\ref{fig:comparison_priority_non-priority_centrists}, as the fraction $s$ raises from $0.02$ to $0.20$. Further details on the emergence of this structure are in Supplementary Material~S3. 

\section{Discussion}
In this paper, we have conducted a comprehensive mechanistic investigation of the impact of a policy change in social media on opinion dynamics. Motivated by Twitter, which recently changed its policy regarding verified users, we proposed a model to simulate the impact of priority users (i.e., users whose content reaches more followers). As a complementary contribution, we also proposed a unified measure for polarization and echo chamber formation, the homophilic bimodality coefficient. Using this framework, we were able to test several scenarios, including cases where stubborn users take advantage of priority accounts. 

In our simulations, we employed homogeneous random (ER) networks as our primary choice because our dynamics is adaptive, resulting in evolving network structures over time. Consequently, all initial configurations characterized by degree correlation, motif patterns, or modular substructures do not significantly influenced the outcome of the dynamics. This notion is supported by a previous systematic test reported in~\cite{de2022modelling}, where the initial network structure was found to have minimal impact on the behavior of the dynamics. Furthermore, in one specific case, depicted in Fig.~\ref{fig:Stubborn_verified}(b), we explored an SF network structure where both the in- and out-degree distributions follow a power distribution. Even in this case, the results were qualitatively similar to those observed in homogeneous random networks. This may occur when the out-degree distribution is concerned, as our dynamics maintain the number of out-edges for each node during rewiring processes. We acknowledge that in future works, exploring different rewiring strategies might lead to different outcomes, where the initial network structure could play a more important role in determining the behavior of the dynamics.

In the first experiment, we tested how the inclusion of priority users could change polarization and echo chamber formation. We showed that when priority users post controversial content, the system remains polarized with echo chambers for a much longer range compared to when the same proportion of those users posts aligned content. Next, we set out to understand what happens when the dynamics admit priority users and those who post aligned content while not changing their opinion (stubborn). Starting from a polarized setting, we saw that for a fixed fraction of priority users, increasing the number of those who are also stubborn amplifies the measurement of echo chambers. Then, we verified that even when the system is not supposed to polarize ($\phi=1.473$), a small fraction of extremists is able to induce polarization and cause echo chambers to form. Importantly, such a transition is accelerated when the extremists become priority users. This is strikingly similar to the change of social norms through reaching a critical mass threshold found in the literature~\cite{centola2018tipping} and constitutes a cautionary tale for unintended emergent effects of policy changes in complex systems with intricate dynamics.

We also considered the effect of centrist users. We found that for a small fraction of centrist ideologues, the echo chambers disappear. However, for a similar scenario but with non-priority stubborn users, with more than $3\%$ of stubborn users, the remaining users become highly polarized. In both cases, the changes in the polarization of opinions $b$ do not imply the formation of echo chambers (as explained in Fig.~\ref{fig:HBC}).

In addition to the effects of priority users on polarization and echo chamber formation, our model also sheds light on the role of ideologues in shaping network dynamics. Regardless of whether users exhibit centrist or extremist behavior, we observe a remarkable phenomenon: when the number of ideologues in the network becomes sufficiently large, the majority of nodes tend to be connected to these ideologues. In other words, a significant portion of the messages exchanged in the network are either sent to or received from these influential users. The latter feature has already been reported for collective human dynamics spurred by Twitter, as in the case of social movements like the 15M in Spain~\cite{borgeholtfoefer2011structural}. Our finding confirms that ideologues play a crucial role in shaping the flow of information and opinions within the social network. Their ability to reach a wide audience allows them to have a significant impact on the formation and reinforcement of echo chambers. When social network algorithms prioritize visibility over content control, these ideologues can use their reach to reinforce their opinions on groups, possibly entrenching echo chamber structures.

Furthermore, our investigation suggests that the influence of these users is not limited to scenarios in which they are committed to extreme opinions. Even centrist ideologues, who may appear as a moderating force on the surface, can have a significant impact on the opinion dynamics when in enough numbers. This implies that addressing the issue of echo chambers and polarization requires a nuanced understanding of the influence exerted by all types of users, regardless of their position on the ideological spectrum.

When put in perspective, the two user behavior mechanisms we investigated, priority users and stubborn users, can be associated to a myriad of diverse behaviors encountered in empirical settings. Prioritizing a user amplifies their influence by means of amplifying their reach. In turn, a committed minority of stubborn users can disturb the system in macroscopic terms. While the stubbornness characteristic can happen in principle by a cause unknown to the platform (for instance, from user self-organization or an externality), prioritization is something we modeled and worked through as a platform's policy. This implies that any policy that boosts user influence should be monitored closely by platforms to ensure the gain of influence is not used maliciously.

In conclusion, our model shows that both priority users and ideologues play critical roles in shaping the dynamics of polarization and echo chamber formation in social media. We acknowledge that all models simplify the explanation of how social networks and opinions behave. However, our findings highlight the need for careful consideration of platform policies that affect the visibility of certain users and the extent of their influence. Therefore, all stakeholders should carefully analyze the possible consequences before changing the policies of social media. By understanding and quantifying the impact of these factors, effective strategies can be developed to mitigate division and promote healthier and more balanced information consumption.

\section{Methods}
\label{sec:methods}
In this section, we present more details about the configuration of the opinion model used and how we analyze the results obtained. We started by setting up the base case (shown in Fig.~\ref{fig:HBC}(a)). We set all the social network users with controversial post contents ($P_p^{\text{con}}$ from Eq.~\ref{eq:P_p_con}) and varied $\phi$ of $P_r$ (Eq.~\ref{eq:rp_I}) between $0$ and $\pi$ (shown in Figure~S1 in the Supplementary Material). The base case was determined by the highest values of $BC$ (Eq.~\ref{eq:BC}) and $BC_{\text{hom}}$ (Eq.~\ref{eq:BC_hom}), with $\phi = 0.785$, which was chosen as it represents the scenario of maximum polarization and echo chamber formation. The parameter $\phi$ in $P_r$ controls the starting point of the cosine-squared function, allowing us to explore the model at different levels of polarization and consensus. Each realization of the model has $10^8$ iterations (which are described in Sec.~\ref{sec:model_description}) to ensure the system reached a nearly stable state, as were all of the following simulations.

The base case test was run on an Erd\H{o}s-R\'enyi~\cite{erdos1960random} network with $n=10^4$ nodes and an edge inclusion probability of $p=1.6 \times 10^{-3}$, resulting in an undirected network with an average degree of 16. The choice of $p$ was based on the desire to create a network with a moderate average degree, which would provide a balance between network connectivity and computational efficiency. Next, to convert this network to directed, we randomly chose the edge direction with equal probability for each possibility. In this way, the resulting network is directed, with an average in-degree and out-degree close to 8. The use of directed networks is crucial for our opinion dynamics model, as it allows for the implementation of the information flow from one user to another through the followership relation.

Furthermore, to create the networks of Fig.~\ref{fig:Stubborn_verified}(b), we use a configuration model~\cite{newman2018networks} whose degree sequence is inverse-sampled from a power law exponent of $\lambda = 2.43$ and a minimum degree of three, meaning the network has an average degree of about 8. In this case, both the degree distributions of the in- and out-edges follow this power law distribution. 

Departing from the base case, we tested the other hypotheses described in this paper. This allows for a comprehensive comparison between each tested configuration and the results of the base case. In order to guarantee that the results obtained are not due to statistical fluctuations, we ran the dynamics 100 times for all results, except for the case of the results presented in Fig.~\ref{fig:comparison_priority_non-priority_centrists}, which were run 500 times. This extensive simulation approach ensures the statistical robustness and reliability of our findings, enabling us to draw meaningful conclusions from the data.

As a complementary result, shown in Fig.~\ref{fig:comparison_priority_non-priority_centrists} of Section~\ref{sec:extremists_advantage}, we tested a different scenario. Specifically, we considered $\phi = 1.473$, a parameter value that leads to a scenario characterized by consensus and no echo chamber formation. In this parametrization, both $BC$ and $BC_{\text{hom}}$ approach their minimum values with low standard deviations, indicating convergence to a stable state. This scenario played a crucial role in our investigation as it allowed us to examine the proportion of extremist ideologues required to induce polarization and form echo chambers. By exploring the impact of different proportions of extremist ideologues in a low-polarization setting, we can understand the conditions under which our model can lead to significant shifts in the overall opinion distribution and the formation of echo chambers.

\section*{Acknowledgments}
The authors would like to thank Carolina Mattsson for fruitful discussions.
Y.M was partially supported by the Government of Aragón, Spain and ``ERDF A way of making Europe'' through grant E36-23R (FENOL), and by Ministerio de Ciencia e Innovación, Agencia Española de Investigación (MCIN/AEI/10.13039/501100011033) Grant No. PID2020-115800GB-I00. We acknowledge the use of the computational resources of COSNET Lab at Institute BIFI, funded by Banco Santander (grant Santander-UZ 2020/0274) and by the Government of Aragón (grant UZ-164255). The funders had no role in study design, data collection and analysis, decision to publish, or preparation of the manuscript.

\section*{Author contributions statement}
H.F.A., K.A.O., and Y.M.: Conceptualization, Methodology, Investigation, Writing - Original Draft, and Writing - Review \& Editing. H.F.A. and K.A.O.: Validation, Formal analysis. H.F.A.: Software.

\section*{Competing interests} 
The authors declare no competing interests.

\bibliography{ref}

\newpage
\section*{Supplementary Material}
\renewcommand{\thefigure}{S\arabic{figure}}
\setcounter{figure}{0}

\renewcommand{\thesection}{S\arabic{section}}
\setcounter{section}{0}
\section{Tuning parameters}
\begin{figure}[!th]
  \centering
    \includegraphics[width=0.499\textwidth]{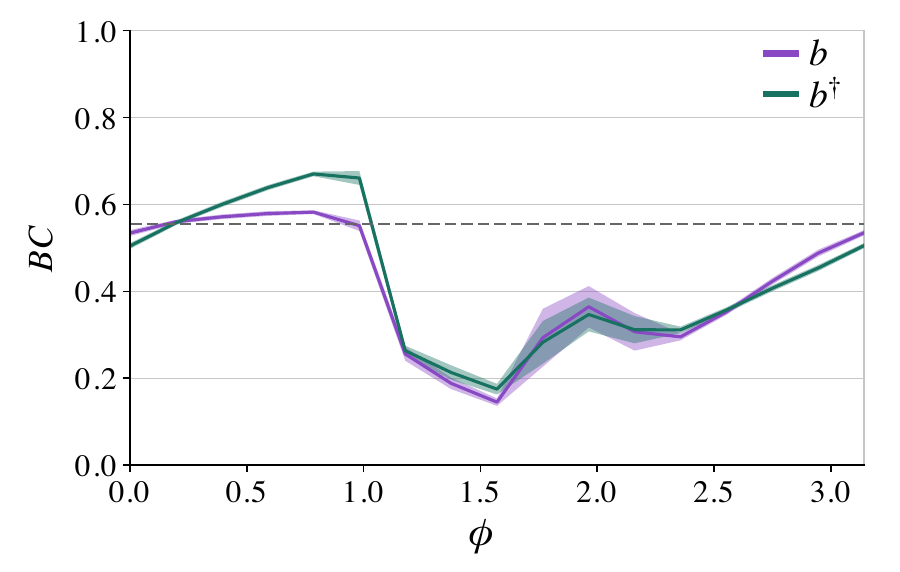}
  \caption{\textbf{Echo chamber and polarization analysis by varying $\phi$:} $BC$ and $BC_{\text{hom}}$ by varying $\phi$. The solid lines represent an average of 100 executions of the dynamics, and the shaded regions are the corresponding standard deviations. Distributions with the bimodality index above the dashed line $BC = 5/9$ can be considered bimodal.}
  \label{fig:BCXphy}
\end{figure}

Figure~\ref{fig:BCXphy}, illustrates the bimodalities, $BC$ and $BC_{\text{hom}}$, computed for the user's opinions $b$ and the diagonal of the map of $b$ versus $b_{NN}$, respectively, according to $\phi$. The results obtained here are different from those found in the previous study~\cite{de2022modelling}. This indicates that the directionality of the network plays an important role in the behavior of the dynamics. 

\section{Comparison between priority and non-priority users}
\begin{figure}[!th]
  \centering
    \includegraphics[width=0.499\textwidth]{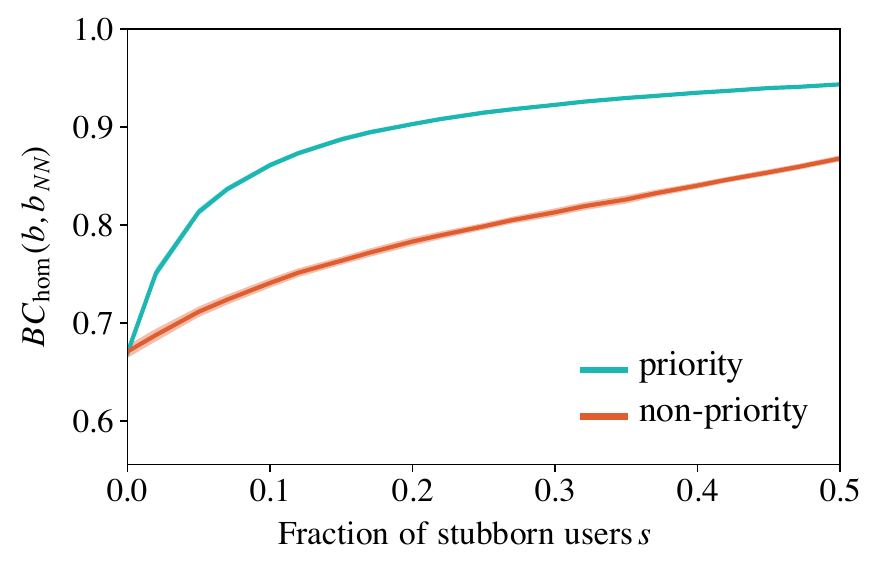}
  \caption{\textbf{Effect of the priority account of stubborn users on the formation of an echo chamber:} Comparison of $BC_{\text{hom}}$ between priority and non-priority users by different percentages of stubborn users. The lines represent an average of over 100 executions of the dynamics, and the shaded regions are the corresponding standard deviations.}
  \label{fig:Stubborn}
\end{figure}

Here we test a different rule for when stubborn users post. That is, they still do not change their opinion but post controversial content this time. Other users still post controversially as well, such as explained in the main text. In this simulation, all of the stubborn users are set with $P^{\text{con}}_p$.
As can be seen in Figure~\ref{fig:Stubborn}, increasing the number of stubborn users strengthens the echo chamber of the network. Echo chamber formation is stronger for a longer range when stubborn users post controversial content rather than aligned.

\begin{figure}[!th]
  \centering
{\includegraphics[width=1\textwidth]{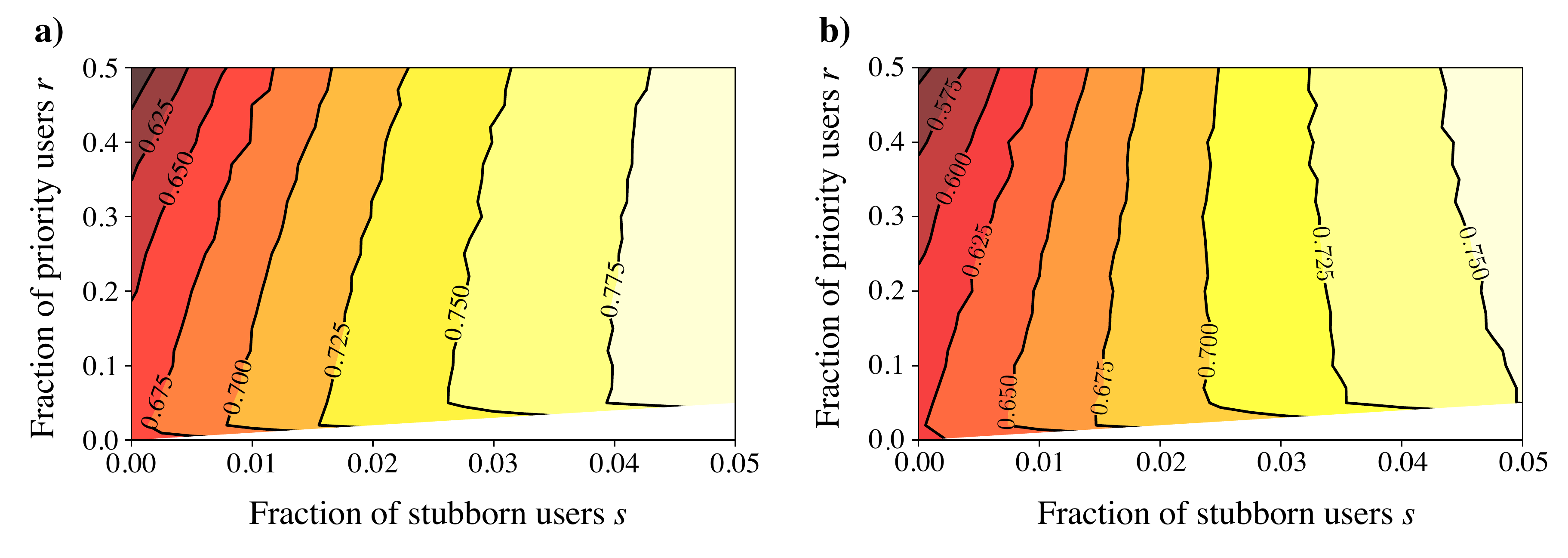}}
  \caption{\textbf{Effect of the relationship between stubborn and priority users:} Contour maps of $BC_{\text{hom}}(\hat{b}, \hat{b}_{NN})$ by varying the percentages of stubborn and verified users. In this case, the stubborn users are not ideologues. While both axes represent fractions of the total number of users $N$, the stubborn users are randomly selected among the priority users (hence the white space for when there are more stubborn than priority users). Panels (a) and (b) represent the dynamics executed on ER and SF networks, respectively. The plots represent an average of 100 runs of the dynamics, and for all tests performed to plot these numbers, the standard deviation was less than 0.02.}
  \label{fig:Stubborn_verified_SM}
\end{figure}

Figure~\ref{fig:Stubborn_verified_SM} maps the relationship between stubborn and priority users. In this test, we do not alter $P_p$. The comparison between ER and SF is relatively similar. In both cases, a small number of stubborn users are able to amplify the echo chamber. Furthermore, the difference in the network structure does not significantly affect the echo chamber structures.

\section{Core-periphery}
\begin{figure}[!th]
  \centering
{\includegraphics[width=1\textwidth]{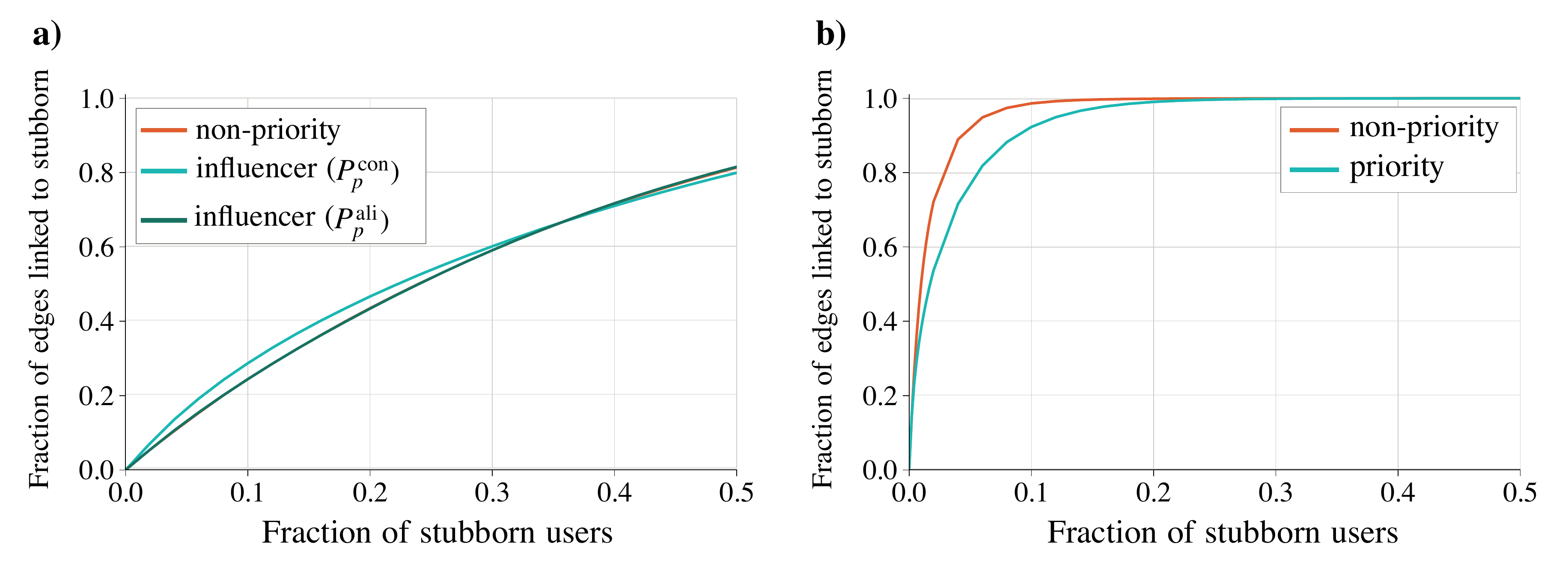}}
  \caption{\textbf{Connectivity of stubborn users:} Comparison between priority and non-priority users regarding the fraction of edges connected to stubborn users. Panels (a) and (b) show the cases of extremist centrist users, respectively. The lines represent an average of over 100 executions of the dynamics, and the shaded regions are the corresponding standard deviations.}
  \label{fig:fraction_of_edges}
\end{figure}

In all scenarios examined, whether stubborn users are centrists or extremists, and whether they are ideologues or not, they tend to be more connected to others in the network. Seemingly, the stubborn users form a core structure that accumulates links within or around, but not outside of it, in a ``hub-and-spoke'' fashion~\cite{gallagher2021coreperiphery}. Figure~\ref{fig:fraction_of_edges} shows that as the number of stubborn users increases, there is a tendency for users to be more connected to them. Figure~\ref{fig:fraction_of_edges}(a) shows the case where the stubborn users are extremists. With and without being ideologues, the fraction of connections to stubborn users tends to be the same. For 50\% of stubborn users, about one out of five connections are in the periphery, not involving stubborn users. In the case of centrists, Figure~\ref{fig:fraction_of_edges}(b), this effect is even clearer. At around 20\% of stubborn users, the proportion of edges in the periphery vanishes, meaning they either connect normal users to stubborn ones or two stubborn users. However, for fractions of stubborn users lower than 0.15, the scenario with non-priority tends to retain stubborn users more connected than in the case with priority users. In both cases, the regular and stubborn users tend to be connected to the stubborn, which leads to the formation of a more connected core, creating the core-periphery structure in the network. We run these tests using ER networks with the same parameters as the previous tests.

\section{Transition into echo chambers}
Here, we re-plot the transition shown in Figure~4 of the main text but using quartiles of the distribution obtained by measuring $BC_hom$ of 500 simulations at each fraction of stubborn users or ideologues.

\begin{figure}[!th]
  \centering
{\includegraphics[width=1\textwidth]{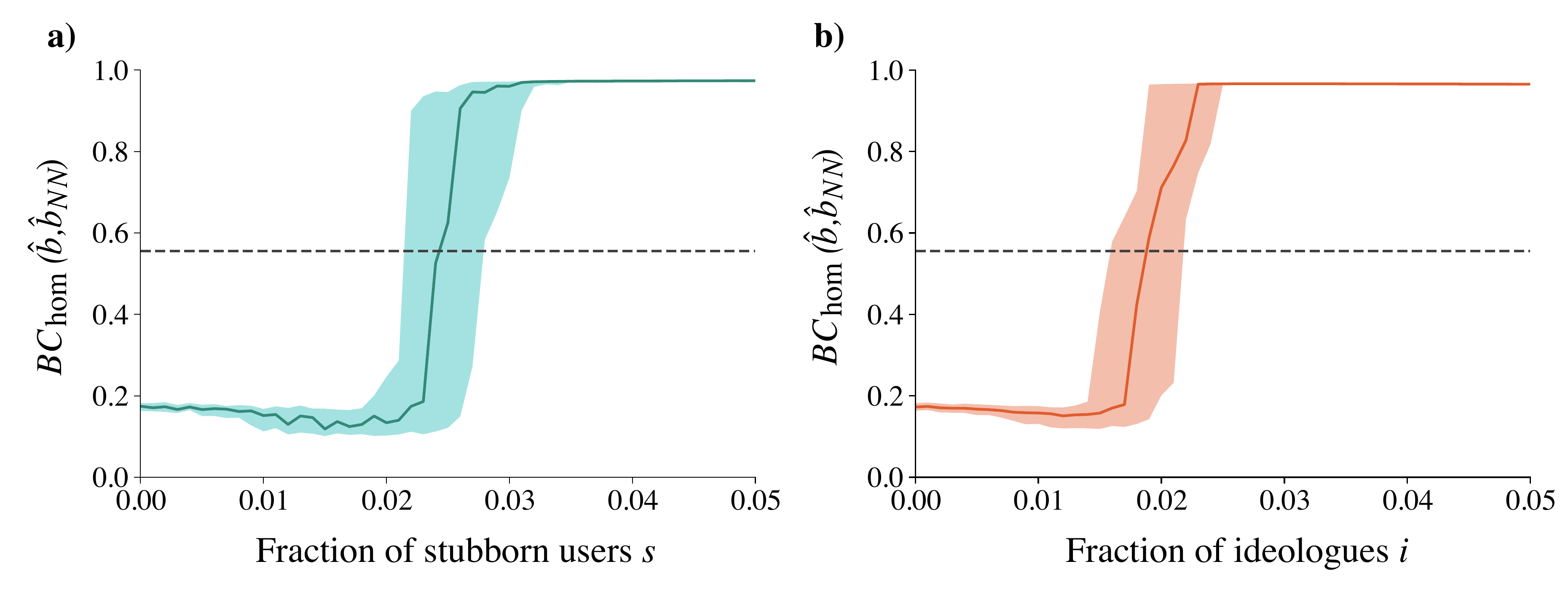}}
  \caption{\textbf{Echo chamber formation by extremist users by quartiles:} the system undergoes a transition as extremist users are added, who are only stubborn in panel (a) and ideologues (stubborn and priority) in panel (b). The bold line represents the median value of the $BC_{\text{hom}}$ calculated from 500 simulations at each value of $s$ and $i$, and the shaded region goes from the first to the third quartile of the same $BC_{\text{hom}}$ distribution.}
  \label{fig:transition}
\end{figure}

\end{document}